\documentclass[conference]{IEEEtran}

\usepackage[T1]{fontenc} 
\usepackage{xcolor}
\usepackage{physics}
\usepackage{amssymb}
\usepackage{algorithm}
\usepackage[noend]{algpseudocode}
\usepackage[font=footnotesize]{caption}
\usepackage[font=footnotesize,subrefformat=parens]{subcaption}
\usepackage{siunitx}

\usepackage[english]{babel} 

\usepackage{booktabs} 
\usepackage{multirow}

\usepackage{enumitem} 
\setlist[itemize]{noitemsep} 

\usepackage[draft]{hyperref} 
\usepackage[capitalize]{cleveref}

\usepackage{cite}

\usepackage[acronyms,nonumberlist,nopostdot,nomain,nogroupskip]{glossaries}
\newacronym{3gpp}{3GPP}{3rd Generation Partnership Project}
\newacronym{adc}{ADC}{Analog to Digital Converter}
\newacronym{5g}{5G}{5th generation}
\newacronym{aimd}{AIMD}{Additive Increase Multiplicative Decrease}
\newacronym{am}{AM}{Acknowledged Mode}
\newacronym{amc}{AMC}{Adaptive Modulation and Coding}
\newacronym[firstplural=Angles of Arrival (AoAs)]{aoa}{AoA}{Angle of Arrival}
\newacronym[firstplural=Angles of Departure (AoDs)]{aod}{AoD}{Angle of Departure}
\newacronym{aqm}{AQM}{Active Queue Management}
\newacronym{awgn}{AGWN}{Additive White Gaussian Noise}
\newacronym{balia}{BALIA}{Balanced Link Adaptation}
\newacronym{bdp}{BDP}{Bandwidth-Delay Product}
\newacronym{bf}{BF}{Beamforming}
\newacronym{cad}{CAD}{Computer Aided Design}
\newacronym{cc}{CC}{Congestion Control}
\newacronym{cdf}{CDF}{Cumulative Distribution Function}
\newacronym{cn}{CN}{Core Network}
\newacronym{cqi}{CQI}{Channel Quality Information}
\newacronym{cp}{CP}{Control Plane}
\newacronym{csirs}{CSI-RS}{Channel State Information - Reference Signal}
\newacronym{dc}{DC}{Dual Connectivity}
\newacronym{dce}{DCE}{Direct Code Execution}
\newacronym{dci}{DCI}{Downlink Control Information}
\newacronym{dl}{DL}{Downlink}
\newacronym{dmr}{DMR}{Deadline Miss Ratio}
\newacronym{dmrs}{DMRS}{DeModulation Reference Signal}
\newacronym{dray}{D-ray}{Deterministic Ray}
\newacronym{e2e}{E2E}{End-to-End}
\newacronym{ecn}{ECN}{Explicit Congestion Notification}
\newacronym{edf}{EDF}{Earliest Deadline First}
\newacronym{enb}{eNB}{evolved Node Base}
\newacronym{epc}{EPC}{Evolved Packet Core}
\newacronym{es}{ES}{Edge Server}
\newacronym{fdma}{FDMA}{Frequency Division Multiple Access}
\newacronym{fdd}{FDD}{Frequency Division Duplexing}
\newacronym[firstplural=Radio Access Technologies (RATs)]{rat}{RAT}{Radio Access Technology}
\newacronym{fov}{FoV}{Field-of-View}
\newacronym{fs}{FS}{Fast Switching}
\newacronym{ftp}{FTP}{File Transfer Protocol}
\newacronym{gnb}{gNB}{Next Generation Node Base}
\newacronym{harq}{HARQ}{Hybrid Automatic Repeat reQuest}
\newacronym{hetnet}{HetNet}{Heterogeneous Network}
\newacronym{hh}{HH}{Hard Handover}
\newacronym{hol}{HOL}{Head-of-Line}
\newacronym{ia}{IA}{Initial Access}
\newacronym{imt}{IMT}{International Mobile Telecommunication}
\newacronym{iot}{IoT}{Internet of Things}
\newacronym{ks}{KS}{Kolmogorov-Smirnov}
\newacronym{los}{LoS}{Line-of-Sight}
\newacronym{lte}{LTE}{Long Term Evolution}
\newacronym{m2m}{M2M}{Machine to Machine}
\newacronym{mac}{MAC}{Medium Access Control}
\newacronym{mc}{MC}{Multi-Connectivity}
\newacronym{mcs}{MCS}{Modulation and Coding Scheme}
\newacronym{mec}{MEC}{Mobile Edge Cloud}
\newacronym{mi}{MI}{Mutual Information}
\newacronym{mimo}{MIMO}{Multiple Input, Multiple Output}
\newacronym{mmwave}{mmWave}{millimeter wave}
\newacronym{mptcp}{MPTCP}{Multipath TCP}
\newacronym{mr}{MR}{Maximum Rate}
\newacronym{mss}{MSS}{Maximum Segment Size}
\newacronym{mtd}{MTD}{Machine-Type Device}
\newacronym{mtu}{MTU}{Maximum Transmission Unit}
\newacronym{nfv}{NFV}{Network Function Virtualization}
\newacronym{nlos}{NLoS}{Non-Line-of-Sight}
\newacronym{nr}{NR}{New Radio}
\newacronym{ofdm}{OFDM}{Orthogonal Frequency Division Multiplexing}
\newacronym{pdcch}{PDCCH}{Physical Downlonk Control Channel}
\newacronym{pdcp}{PDCP}{Packet Data Convergence Protocol}
\newacronym{pdsch}{PDSCH}{Physical Downlink Shared Channel}
\newacronym{pdu}{PDU}{Packet Data Unit}
\newacronym{pf}{PF}{Proportional Fair}
\newacronym{pgw}{PGW}{Packet Gateway}
\newacronym{phy}{PHY}{Physical}
\newacronym{pbch}{PBCH}{Physical Broadcast Channel}
\newacronym[plural=\gls{mme}s,firstplural=Mobility Management Entities (MMEs)]{mme}{MME}{Mobility Management Entity}
\newacronym{prb}{PRB}{Physical Resource Block}
\newacronym{pss}{PSS}{Primary Synchronization Signal}
\newacronym{pucch}{PUCCH}{Physical Uplink Control Channel}
\newacronym{pusch}{PUSCH}{Physical Uplink Shared Channel}
\newacronym{qd}{QD}{Quasi-Deterministic}
\newacronym{rach}{RACH}{Random Access Channel}
\newacronym{ran}{RAN}{Radio Access Network}
\newacronym{red}{RED}{Random Early Detection}
\newacronym{rf}{RF}{Radio Frequency}
\newacronym{rlc}{RLC}{Radio Link Control}
\newacronym{rlf}{RLF}{Radio Link Failure}
\newacronym{rrc}{RRC}{Radio Resource Control}
\newacronym{rrm}{RRM}{Radio Resource Management}
\newacronym{rr}{RR}{Round Robin}
\newacronym{rs}{RS}{Remote Server}
\newacronym{rsrp}{RSRP}{Reference Signal Received Power}
\newacronym{rss}{RSS}{Received Signal Strength}
\newacronym{rt}{RT}{Ray-Tracer}
\newacronym{rtt}{RTT}{Round Trip Time}
\newacronym{rw}{RW}{Receive Window}
\newacronym{rx}{RX}{Receiver}
\newacronym{sa}{SA}{standalone}
\newacronym{sack}{SACK}{Selective Acknowledgment}
\newacronym{sap}{SAP}{Service Access Point}
\newacronym{sch}{SCH}{Secondary Cell Handover}
\newacronym{scoot}{SCOOT}{Split Cycle Offset Optimization Technique}
\newacronym{sdma}{SDMA}{Spatial Division Multiple Access}
\newacronym{sinr}{SINR}{Signal to Interference plus Noise Ratio}
\newacronym{sm}{SM}{Saturation Mode}
\newacronym{snr}{SNR}{Signal-to-Noise-Ratio}
\newacronym{son}{SON}{Self-Organizing Network}
\newacronym{ss}{SS}{Synchronization Signal}
\newacronym{srs}{SRS}{Sounding Reference Signal}
\newacronym{sss}{SSS}{Secondary Synchronization Signal}
\newacronym{tb}{TB}{Transport Block}
\newacronym{tcp}{TCP}{Transmission Control Protocol}
\newacronym{tdd}{TDD}{Time Division Duplexing}
\newacronym{tdma}{TDMA}{Time Division Multiple Access}
\newacronym{tfl}{TfL}{Transport for London}
\newacronym{tm}{TM}{Transparent Mode}
\newacronym{trp}{TRP}{Transmitter Receiver Pair}
\newacronym{tti}{TTI}{Transmission Time Interval}
\newacronym{ttt}{TTT}{Time-to-Trigger}
\newacronym{tx}{TX}{Transmitter}
\newacronym{ue}{UE}{User Equipment}
\newacronym{ul}{UL}{Uplink}
\newacronym{uml}{UML}{Unified Modeling Language}
\newacronym{um}{UM}{Unacknowledged Mode}
\newacronym{utc}{UTC}{Urban Traffic Control}
\newacronym{vm}{VM}{Virtual Machine}
\newacronym{rsrq}{RSRQ}{Reference Signal Received Quality}
\newacronym{rssi}{RSSI}{Received Signal Strength Indicator}
\newacronym{crs}{CRS}{Cell Reference Signal}
\newacronym{nsa}{NSA}{Non Stand Alone}
\newacronym{mrdc}{MR-DC}{Multi \gls{rat} \gls{dc}}
\newacronym{endc}{EN-DC}{E-UTRAN-\gls{nr} \gls{dc}}
\newacronym{5gc}{5GC}{5G Core}
\newacronym{si}{SI}{Study Item}
\newacronym{iab}{IAB}{Integrated Access and Backhaul}
\newacronym{wf}{WF}{Wired-first}
\newacronym{hqf}{HQF}{Highest-quality-first}
\newacronym{pa}{PA}{Position-aware}
\newacronym{mlr}{MLR}{Maximum-local-rate}
\newacronym{wbf}{WBF}{Wired Bias Function}
\newacronym{mib}{MIB}{Master Information Block}
\newacronym{sib}{SIB}{Secondary Information Block}
\newacronym{kpi}{KPI}{Key Performance Indicator}
\newacronym{ppp}{PPP}{Poisson Point Process}
\newacronym{mpc}{MPC}{Multi Path Component}
\newacronym{wlan}{WLAN}{Wireless Local Area Network}
\newacronym{scm}{SCM}{Spatial Channel Model}

\usepackage{tikz}
\usepackage{pgfplots}
\newlength\fheight
\newlength\fwidth
\pgfplotsset{compat=newest} 
\pgfplotsset{plot coordinates/math parser=false}
\usetikzlibrary{plotmarks,patterns,decorations.pathreplacing,backgrounds,calc,arrows,arrows.meta,spy,matrix}
\usepgfplotslibrary{patchplots,groupplots}
\usepackage{tikzscale}
\tikzset{>=latex}

\IEEEoverridecommandlockouts
\newcommand\copyrighttext{%
  \footnotesize \textcopyright 2020 IEEE. Personal use of this material is permitted.
  Permission from IEEE must be obtained for all other uses, in any current or future media, including reprinting/republishing this material for advertising or promotional purposes, creating new collective works, for resale or redistribution to servers or lists, or reuse of any copyrighted component of this work in other works.}
\newcommand\copyrightnotice{%
\begin{tikzpicture}[remember picture,overlay]
\node[anchor=south,yshift=5pt] at (current page.south) {\fbox{\parbox{\dimexpr\textwidth-\fboxsep-\fboxrule\relax}{\copyrighttext}}};
\end{tikzpicture}%
}

\def\BibTeX{{\rm B\kern-.05em{\sc i\kern-.025em b}\kern-.08em
    T\kern-.1667em\lower.7ex\hbox{E}\kern-.125emX}}

\newcommand{\normalDistrib}[2]{\ensuremath{\mathcal{N} \qty(#1, #2)}}
\newcommand{\ricianDistrib}[2]{\ensuremath{\mathcal{R} \qty(#1, #2)}} 
\newcommand{\laplacianDistrib}[2]{\ensuremath{\mathcal{L} \qty(#1, #2)}}
\newcommand{\exponentialDistrib}[1]{\ensuremath{\mathcal{E} \qty(#1)}}

\title{Quasi-Deterministic Channel Model for mmWaves:\\Mathematical Formalization and Validation} 
\author{\IEEEauthorblockN{Mattia Lecci\IEEEauthorrefmark{1}, 
                          Michele Polese\IEEEauthorrefmark{3},
                          Chiehping Lai\IEEEauthorrefmark{2},
                          Jian Wang\IEEEauthorrefmark{2},
                          Camillo Gentile\IEEEauthorrefmark{2},
                          Nada Golmie\IEEEauthorrefmark{2},
                          Michele Zorzi\IEEEauthorrefmark{1}}\\
        \IEEEauthorblockA{\vspace{0cm}    \small \IEEEauthorrefmark{1}Department of Information Engineering, University of Padova, Italy, e-mail: \texttt{\{name.surname\}@dei.unipd.it}\\
        \small \IEEEauthorrefmark{3}Institute for the Wireless Internet of Things, Northeastern University, Boston, MA, USA, e-mail: \texttt{m.polese@northeastern.edu}\\
        \small \IEEEauthorrefmark{2}National Institute of Standards and Technology (NIST), Gaithersburg, MD, USA, e-mail: \texttt{\{name.surname\}@nist.gov}}
        \thanks{This work was partially supported by NIST under Award No. 70NANB18H273.
        Mattia Lecci's activities were supported by \textit{Fondazione CaRiPaRo} under the grants ``Dottorati di Ricerca'' 2018.}
}

\IEEEoverridecommandlockouts 
\begin{document}

\makeatletter
\patchcmd{\@maketitle}
  {\addvspace{0.5\baselineskip}\egroup}
  {\addvspace{-1\baselineskip}\egroup}
  {}
  {}
\makeatother

\flushbottom
\setlength{\parskip}{0ex plus0.1ex}

\maketitle
\copyrightnotice

\begin{abstract}
5G and beyond networks will use, for the first time ever, the \gls{mmwave} spectrum for mobile communications.
Accurate performance evaluation is fundamental to the design of reliable \gls{mmwave} networks, with accuracy rooted in the fidelity of the channel models.
At \glspl{mmwave}, the model must account for the spatial characteristics of propagation since networks will employ highly directional antennas to counter the much greater pathloss.
In this regard, \gls{qd} models are highly accurate channel models, which characterize the propagation in terms of clusters of multipath components, given by a reflected ray and multiple diffuse components of any given \gls{cad} scenario.
This paper introduces a detailed mathematical formulation for \gls{qd} models at \glspl{mmwave}, that can be used as a reference for their implementation and development.
Moreover, it compares channel instances obtained with an open source NIST \gls{qd} model implementation against real measurements at 60~GHz, substantiating the accuracy of the model.
Results show that, when comparing the proposed model and deterministic rays alone with a measurement campaign, the \gls{ks} test of the QD model improves by up to 0.537.
\end{abstract}

\begin{tikzpicture}[remember picture,overlay]
\node[anchor=north,yshift=-10pt] at (current page.north) {\parbox{\dimexpr\textwidth-\fboxsep-\fboxrule\relax}{\centering \footnotesize This paper has been presented at IEEE GLOBECOM 2020. \textcopyright 2020 IEEE.\\
  Please cite it as: M. Lecci, M. Polese, C. Lai, J. Wang, C. Gentile, N. Golmie, M. Zorzi, ``Quasi-Deterministic Channel Model for mmWaves: Mathematical Formalization and Validation,'' IEEE Global Communications Conference (GLOBECOM), Dec. 2020, Taipei, Taiwan}};
\end{tikzpicture}%

\glsresetall



\section{Introduction}

To satisfy a constantly growing demand for mobile connectivity,
the future generations of wireless networks will exploit frequencies above 6~GHz for radio access.
This portion of the spectrum, loosely identified as the \gls{mmwave} band, features large chunks of untapped spectrum, to be used to provide ultra-high datarates to end users.
In this regard, cellular networks implementing the 3GPP NR Release 15 specifications can support a carrier frequency of up to 52.6~GHz, while IEEE 802.11ad/ay foresee \glspl{wlan} operating in the unlicensed spectrum at 60~GHz.
The development of robust mobile networks in this frequency range is challenging.
The high propagation loss, indeed, limits the coverage of the mmWave base stations and access points.
Besides, \gls{mmwave} signals are blocked by common obstacles (e.g., the human body, walls, vehicles) with high penetration losses, making the power of the received signal highly variable.

A reliable and accurate evaluation of the performance is fundamental to the development of technological solutions for \gls{mmwave} cellular networks.
Given the difficulties associated to a testbed setup at such high frequencies, the research community has, so far, mostly relied on analysis and simulations~\cite{bai2015coverage}, developing several tools for different protocol stacks and proposed communication technologies~\cite{mezzavilla2018endtoend,assasa2019implementation,patriciello2019e2e}.
The accuracy of the performance evaluation, however, depends to a large degree on the fidelity of the representation of the channel~\cite{polese2018impact}.
When it comes to \glspl{mmwave}, the complex dynamics of the propagation environment strengthen the need for a comprehensive model, which accounts not only for the pathloss, but also for the spatial behavior of the signal propagation and its interaction with directional antennas, and for the fading that arises from the interaction with the scatterers in the environment~\cite{rangan2017potentials}.

This need has sparked several research efforts aimed at characterizing the \gls{mmwave} channel.
Measurement campaigns have been conducted in diverse settings, e.g., in urban or rural scenarios~\cite{rappaport2013millimeter,hur2016proposal}, or indoors~\cite{gentile2018qdDataCenter,lai2019methodology}.
These works have identified a number of key elements for the modeling of \gls{mmwave} channels~\cite{hemadeh2018millimeter,rappaport2017overview}: (i) the multipath components are sparse in the angular domain, and this impacts the characterization and design of beamforming schemes; (ii) blockage affects the link dynamics much more than at sub-6~GHz; (iii) effects of diffuse scattering from rough surfaces become more prominent at shorter wavelengths.
So far, different modeling approaches have emerged in the \gls{mmwave} domain.
The simplest ones are used, generally, for mathematical analysis, and characterize fading with Nakagami-m or Rayleigh random variables, often with simplified beamforming patterns~\cite{bai2015coverage}.
The \gls{3gpp} has adopted a \gls{scm} for the evaluation of NR in the frequency range between 0.5 and 100~GHz, in which a channel matrix is generated with a purely stochastic approach~\cite{3gpp.38.901}.
These approaches, however, cannot fully capture the fading and angular components of the \gls{mmwave} channel that relate with a \textit{realistic} and \textit{specific} propagation environment.

This can be achieved using a \gls{rt}~\cite{degliesposti2014rt}, which models the channel by generating the \glspl{mpc} that, given the description of a certain scenario, can physically propagate from the transmitter's to the receiver's location.
These \glspl{mpc} are characterized by angles of arrival and departure, power and delay, and can either be the direct component, or rays reflected from the scattering surfaces of the environment~\cite{lai2019methodology}.
Additionally, a \gls{rt}, which is purely deterministic and only depends on the geometry of the scenario, can be combined with stochastic models for the generation of diffuse components to create a \textit{\gls{qd}} model.
These depend on the roughness of the surface on which rays reflect, and are clustered around the main reflected component~\cite{maltsev2016channel}.
The modeling of these components is relevant at \glspl{mmwave} as the wavelength approaches the scale of the surface roughness~\cite{gentile2018qdDataCenter}.

\gls{qd} models for \glspl{mmwave} have been introduced in~\cite{maltsev2016channel,lai2019methodology}.
These papers, however, discuss the measurement process and the derivation of the parameters for the model, but only give a high-level overview of the mathematical formulation of the \gls{qd} model.
The goal of this paper is to fill that void, namely to provide the \gls{mmwave} research community a detailed recipe on how to generate realizations of NIST's implementation of the IEEE 802.11ay \gls{qd} model.
We will discuss the generation of a channel instance step by step, precisely describing the parameters and random distributions, using an open source \gls{qd} implementation developed by NIST and the University of Padova as a reference\footnote{Available at \url{https://github.com/signetlabdei/qd-realization/tree/feature/treetraversal}.}.
Additionally, we will compare channels generated using this \gls{qd} model with real measurements in an indoor environment at 60~GHz, to validate the accuracy of the model.

The rest of this paper is organized as follows.
\cref{sec:notation} introduces the notation that will be used throughout the paper.
\cref{sec:model} reports the mathematical model, with the comparison in \cref{sec:comparison_with_measurements}.
Finally, \cref{sec:conclusions} concludes the paper.

\section{Notation}
\label{sec:notation}

In the remainder of this paper, simple math font (e.g., $a$) is used for both scalar and vector variables, while bold math font is used for random variables (e.g., $\vb*{a}$).
The function $d(x_1, x_2)$ corresponds to the euclidean distance between points $x_1$ and $x_2$ in 3D space.
The following notation and distributions for random variables are assumed:
\begin{itemize}
  \item $\vb{X} \sim \normalDistrib{\mu}{\sigma^2}$: Normal distribution with $\mathbb{E}[\vb{X}] = \mu$ and $\mathrm{var}(\vb{X}) = \sigma^2$
  \item $\vb{X} \sim \ricianDistrib{s}{\sigma}$: Rician distribution where $s,\sigma \geq 0$.
  It can be generated as $\vb{X} = \sqrt{\vb{Y} + \vb{Z}}$, where $\vb{Y} \sim \normalDistrib{s}{\sigma^2}$, $\vb{Z} \sim \normalDistrib{0}{\sigma^2}$.
  \item $\vb{X} \sim \laplacianDistrib{\mu}{\sigma^2}$: Laplacian distribution with $\mathbb{E}[\vb{X}] = \mu$ and $\mathrm{var}(\vb{X}) = \sigma^2$
  \item $\vb{X} \sim \exponentialDistrib{\lambda}$: Exponential distribution with $\mathbb{E}[\vb{X}] = \frac{1}{\lambda}$ and $\mathrm{var}(\vb{X}) = \frac{1}{\lambda^2}$
  \item $\vb{X} \sim \mathcal{U}[a,b]$: Uniform distribution in the closed interval $[a,b]$
\end{itemize}

\section{Mathematical Model}
\label{sec:model}

In this section, we will provide a step-by-step tutorial on how to generate a channel with a \gls{qd} model, with a precise and rigorous mathematical formulation.

The \gls{qd} model considers as a basis a deterministic channel, which can be computed through ray tracing for time $t$, given an environment geometry, and \gls{tx} and \gls{rx} positions~\cite{degliesposti2014rt}.
The computed \glspl{dray} will then be the baseline for the multipath components randomly generated by the \gls{qd} model.
If present, the direct ray is treated separately as it does not generate any diffuse component.

The \gls{qd} model can be realized from the model for a first-order reflection and from it generalized to higher-order reflections.
For reasons that will become clear later on, we define the instant in which the direct ray should arrive at the \gls{rx} (even if it is actually blocked) as $t_0 = t + t_{\rm dir}$, where $t_{\rm dir} = \frac{d(\mathrm{\gls{tx}}, \mathrm{\gls{rx}})}{c}$, and $c$ is the speed of light.
From now on we will consider a frame of reference in the variable $\tau$ relative to time $t_0$, where $\tau = 0$ corresponds to $t_0$.
Given this choice, the direct ray, if it exists, will arrive at time $\tau = 0$, whereas the reflected \glspl{dray} will arrive at times $\tau > 0$.

\subsection{First-order reflections}
\label{sub:first_order_reflections}

Statistics for all rays are assumed independent of their arrival time.
We thus consider, without loss of generality, a single reflected \gls{dray} with arrival time $\tau_0 > 0$, path gain $PG_0$, \gls{aod} along the azimuth/elevation axes $AoD_{az/el, 0}$, and \gls{aoa} $AoA_{az/el, 0}$.
The same procedure will be repeated for all other reflected \glspl{dray}.

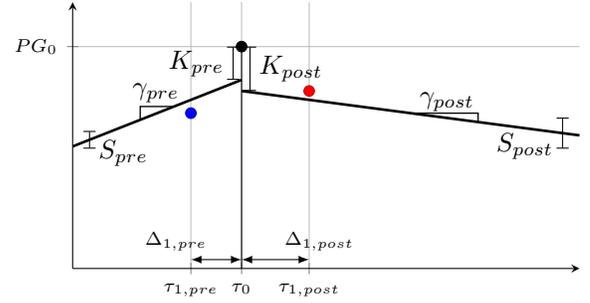
\begin{figure}[t]
  \centering
  \setlength\belowcaptionskip{-.6cm}
  \setlength\fwidth{0.8\columnwidth} 
  \setlength\fheight{0.4\columnwidth}
%
%
\begin{tikzpicture}
\pgfplotsset{every tick label/.append style={font=\scriptsize}}

\begin{axis}[%
width=0.951\fwidth,
height=\fheight,
at={(0\fwidth,0\fheight)},
scale only axis,
xmin=-1,
xmax=2,
xtick={-0.3, 0, 0.4},
xticklabels={{$\tau_{1, pre}$}, {$\tau_0$}, {$\tau_{1, post}$}},
ymin=0,
ymax=12,
ytick={10},
yticklabels={{$PG_0$}},
axis background/.style={fill=white},
axis lines=left, 
xmajorgrids,
ymajorgrids,
legend style={legend cell align=left, align=left, draw=white!15!black}
]
\addplot[ycomb, color=black, mark=*, mark options={solid, fill=black, black}, forget plot] table[row sep=crcr] {%
0	10\\
};

\draw [|-|] (-0.05, 10) -- (-0.05, 8.5);
\node[left, align=right] at (axis cs:-0.05,9.25) {$K_{pre}$};

\addplot [color=black, line width=1pt]
  table[row sep=crcr]{%
0 8.5\\
-1  5.5\\
};

\addplot [color=black, line width=0.5pt]
  table[row sep=crcr]{%
-0.4 7.3\\
-0.6 7.3\\
-0.6 6.7\\
};
\node[right, align=left] at (axis cs:-0.7,7.9) {$\gamma_{pre}$};

\draw [|-|] (-0.9, 5.4) -- (-0.9, 6.2);
\node[right, align=left] at (axis cs:-0.9, 5.2) {$S_{pre}$};

\draw [|-|] (0.05, 10) -- (0.05, 8);
\node[right, align=left] at (axis cs:0.05,9) {$K_{post}$};

\addplot [color=black, line width=1pt]
  table[row sep=crcr]{%
0	8\\
2	6\\
};

\addplot [color=black, line width=0.5pt]
  table[row sep=crcr]{%
1 7\\
1.4 7\\
1.4 6.6\\
};
\node[right, align=left] at (axis cs:1,7.5) {$\gamma_{post}$};

\draw [|-|] (1.9, 5.4) -- (1.9, 6.8);
\node[left, align=right] at (axis cs:1.9, 5.5) {$S_{post}$};

\addplot[scatter, only marks, color=blue, mark=*, mark options={solid, fill=blue, blue}, forget plot] table[row sep=crcr] {%
0.4 8\\
-0.3 7\\
};

\draw [<->] (0, 0.4) -- (0.4, 0.4);
\node [above right, align=left] at (axis cs:0.2, 0.4) {\scriptsize $\Delta_{1, post}$};

\draw [<->] (0, 0.4) -- (-0.3, 0.4);
\node [above left, align=right] at (axis cs: -0.15, 0.4) {\scriptsize $\Delta_{1, pre}$};

\end{axis}
\end{tikzpicture}%
  \caption{Graphical representation of \gls{qd} parameters.}
  \label{fig:qdFig}
\end{figure}

\begin{algorithm}[t] \caption{Single Reflection QD Generator} \label{alg:singleReflectionQdGenerator}
\begin{algorithmic}[1]
\small
\Function{GetMpcsFirstReflection}{Cursor: $\tau_0$, $PG_{0,D}$, $AoD_{az/el,0}$, $AoA_{az/el,0}$, Material}
    \State $RL \leftarrow \ricianDistrib{s_{RL,Material}}{\sigma_{RL,Material}}$
    \State $PG_0 = PG_{0,D} - (RL - \mu_{RL})$
    
    \vspace*{1ex}
    
    \State PreCursors $\leftarrow$ \Call{ComputePre/PostCursors}{$\tau_0$, $PG_0$, $AoD/AoA_{az/el,0}$, Material}
    \State PostCursors $\leftarrow$ \Call{ComputePre/PostCursors}{$\tau_0$, $PG_0$, $AoD/AoA_{az/el,0}$, Material}
    
    \vspace*{1ex}
    
    \Return PreCursors, Cursor, PostCursors
\EndFunction

\vspace*{2ex}

\Function{ComputePre/PostCursors}{$\tau_0$, $PG_0$, $AoD/AoA_{az/el,0}$, Material}
    \State $\lambda \leftarrow \ricianDistrib{s_{\lambda,Material}}{\sigma_{\lambda,Material}}$
    \State $\Delta_i \leftarrow \exponentialDistrib{\lambda}, \quad i = 1,\ldots,N_{pre/post}$
    \State $\tau_i = \tau_0 \pm \sum_{j=1}^i \Delta_i$ \Comment{\textit{Add for post-cursors, subtract for pre-cursors}}
    \State Remove pre-cursors with $\tau_i < 0$, update $N_{pre/post}$
    
    \vspace*{1ex}
    
    \State $K_{dB} \leftarrow \ricianDistrib{s_{K,Material}}{\sigma_{K,Material}}$
    \State $\gamma \leftarrow \ricianDistrib{s_{\gamma,Material}}{\sigma_{\gamma, materia}}$
    \State $\sigma_{s,Material} \leftarrow \ricianDistrib{s_{\sigma_{s,Material}}}{\sigma_{\sigma_{s,Material}}}$
    \State $S_i \leftarrow \normalDistrib{0}{\sigma_{s,Material}^2}$
    \State $PG_i = PG_{0,dB} - K_{dB} - 10\log_{10}(e)\, \frac{|\tau_i - \tau_0|}{\gamma} + 10\log_{10}(e)\, S_i$
    \State Remove \glspl{mpc} with $PG_i \geq PG_0$, update $N_{pre/post}$
    
    \vspace*{1ex}
    
    \State $\sigma_\alpha \leftarrow \ricianDistrib{\mu_{\sigma_\alpha}}{\sigma_{\sigma_\alpha}}$
    \State $\alpha_{AoD/AoA, az/el,i} \leftarrow \laplacianDistrib{0}{\sigma_\alpha^2}$
    \State $AoD/AoA_{az/el,i} \leftarrow AoD/AoA_{az/el,0} + \alpha_{AoD/AoA,az/el,i}$
    \State Wrap angles in $az = [0, 360)$, $el = [0, 180]$
    
    \vspace*{1ex}
    
    \State $\phi_i \leftarrow \mathcal{U}[0, 2\pi)$
    
    \vspace*{1ex}
    
    \Return ($\tau_i$, $PG_i$, $AoD/AoA_{az/el,i}$)
\EndFunction

\end{algorithmic}
\end{algorithm}

A cluster can be defined as the set with a \gls{dray} and the corresponding \glspl{mpc}.
The total number of \glspl{mpc} of a given cluster will be $N_{\gls{mpc}} = N_{pre} + 1 + N_{post}$, including pre-cursors (i.e., diffuse components that are received before the \gls{dray}), main cursor (i.e., the \gls{dray}), and post-cursors (i.e., received after the \gls{dray}).
Based on some experimental evidence, we suggest to use $N_{pre} = 3$ and $N_{post} = 16$, although these numbers may vary in different locations and models.

The arrival times of the \glspl{mpc} are modeled as a Poisson process, meaning that their inter-arrival times are independent and exponentially distributed.
Namely, the post-cursors arrival times $\vb*{\tau}_{i, post}$ are random variables generated based on inter-arrival delays $\vb*{\Delta}_{i, post} = \vb*{\tau}_{i, post} - \vb*{\tau}_{i-1, post}$ as follows
\begin{equation}
  \vb*{\Delta}_{i, post} | \vb*{\tau}_{i-1} \sim \exponentialDistrib{\vb*{\lambda}_{post}},
\end{equation}
for $i = 1, \ldots, N_{post}$, where the arrival rate $\vb*{\lambda}_{post} \sim \ricianDistrib{s_{\lambda_{post}}}{\sigma_{\lambda_{post}}}$ is a random variable itself.
With slight abuse of notation, we consider $\vb*{\tau}_{0, post} = \tau_0$, i.e., the time of arrival of the \gls{dray}.
Post-cursors arrival times are then computed as 
\begin{equation}\label{eq:tau_post}
  \vb*{\tau}_{i, post} = \vb*{\tau}_{i-1, post} + \vb*{\Delta}_{i, post} = \tau_0 + \sum_{j=1}^{i} \vb*{\Delta}_{j, post},
\end{equation}
 for $i = 1,\, \ldots,\, N_{post}$.
Please note that random parameters such as $\lambda_{post}$ should be extracted independently for each \gls{dray}.

Pre-cursors will be similarly generated, with the difference that \cref{eq:tau_post} will subtract inter-arrival delay, thus making $\vb*{\tau}_{i, pre} < \tau_0$ for $i=1,\, \ldots,\, N_{pre}$.

Since the number of pre/post-cursors was empirically extrapolated from measured data from~\cite{lai2019methodology}, during the \gls{qd} model generation some of them may not follow some basic assumptions.
For example, when a \gls{dray} has a delay $\tau_0$ close to $0$, some of its generated pre-cursors might arrive before the direct ray itself.
Since this situation cannot happen in the physical reality, rays with $\tau_{i, pre} < 0$ are removed and $N_{pre}$ is consequently updated.

The path gain of the \gls{dray} is
\begin{equation}
  \vb{PG}_{0} = 20 \log_{10}\qty(\frac{\lambda_c}{4\pi \ell_{ray}}) - \vb{RL}_{dB},
\end{equation}
where $\lambda_c$ is the wavelength of the carrier frequency, $\ell_{ray}$ is the total ray length, and $\vb{RL} \sim \ricianDistrib{s_{RL}}{\sigma_{RL}}$ is the random reflection loss factor given by the reflecting surface's material.
If only the deterministic part of the ray-tracer is considered, the path gain $PG_{0,D}$ only includes the mean reflection loss $\mu_{RL}$.

Once the arrival times $\vb*{\tau}_i$ are known, the path gains for the \glspl{mpc} can be computed as
\begin{equation}\label{eq:pg_db}
\begin{split}
  \vb{PG}_{pre/post, i, \mathrm{dB}} ={} & \vb*{PG}_{0, \mathrm{dB}} - \vb*{K}_{pre/post, \mathrm{dB}} + \\
                            & -\frac{|\vb*{\tau}_{i, pre/post} - \tau_0|}{\vb*{\gamma}_{pre/post}} (10\log_{10}e) +\\
                            &  (10\log_{10}e) \vb{S}_{pre/post} ,
\end{split}
\end{equation}
where
\begin{itemize}
  \item $\vb{K}_{pre/post, \mathrm{dB}} \sim \ricianDistrib{s_{K_{pre/post}}}{\sigma_{K_{pre/post}}}$ is a loss factor,
  \item $\vb*{\gamma}_{pre/post} \sim \ricianDistrib{s_{\gamma_{pre/post}}}{\sigma_{\gamma_{pre/post}}}$ is the power-delay decay constant,
  \item $\vb{S}_{pre/post} \sim \normalDistrib{0}{\vb*{\sigma}_{s, pre/post}^2}$ is the power-delay decay standard deviation, where $\vb*{\sigma}_{s, pre/post} \sim \ricianDistrib{s_{\sigma_{s,pre/post}}}{\sigma_{\sigma_{s,pre/post}}}$.
\end{itemize}
While $\vb{K}_{pre/post, \mathrm{dB}}$, $\vb*{\gamma}_{pre/post}$, and $\vb*{\sigma}_{s, pre/post}$ are independent across clusters, and $\vb{S}_{pre/post}$ is independently extracted for each \gls{mpc}.

Since the main cursor is that with the maximum $PG$ when extracting the statistics from the measurements, \glspl{mpc} with $\vb{PG}_{pre/post, i} \geq PG_{0,D}$ are removed, updating, in this case, $N_{pre/post}$.

Finally, the angle of departure in azimuth (and similarly the \gls{aod} in elevation and the \glspl{aoa} in azimuth and elevation) of the \glspl{mpc} are computed as
\begin{equation}
  \vb{AoD}_{az, i} = AoD_{az,0} + \vb*{\alpha}_{AoD,az,i},
\end{equation}
where $\vb*{\alpha}_{AoD,az,i} \sim \laplacianDistrib{0}{\vb*{\sigma}_{\alpha_{AoD,az}}^2}$ is the angle spread.
The variance $\vb*{\sigma}_{\alpha_{AoD,az}}^2 \sim \ricianDistrib{s_{\sigma_{\alpha_{AoD,az}}^2 }}{\sigma_{\sigma_{\alpha_{AoD,az}}^2 }}$ is itself a random variable independently extracted for each cluster.

Finally, the phase shift $\vb*{\phi}_i$ due to both diffusion and Doppler shift is considered $\mathcal{U}[0,2\pi)$ independently for each diffuse \gls{mpc}.

\subsection{Higher-order reflections}
\label{sub:higher_order_reflections}
For the $n^{th}$ reflection order, with $n >1$, multiple heuristics can be thought of to compute the diffuse components.
Unfortunately, the measurements taken and the models adopted to process them do not allow for a reliable confirmation of the proposed heuristics, but an extension to higher reflection orders is nevertheless needed for inclusion in a generic ray-tracer.

The path gain for specular rays with $n$ reflections is extended as follows:
\begin{equation}
  \vb{PG}_{0} = 20\log_{10} \qty(\frac{\lambda_c}{4\pi \ell_{ray}}) - \sum_{i=1}^n \vb{RL}_{i, dB},
\end{equation}
where $\vb*{RL}_{i, dB} \sim \ricianDistrib{s_{RL,i}}{\sigma_{RL,i}}$, and $(s_{RL,i}, \sigma_{RL,i})$ refers to the statistics associated to the material of the $i$-th reflector of the given ray.

We propose two simple heuristics: a complete multiple reflection \gls{qd} model and a reduced multiple reflection \gls{qd} model.

\paragraph*{Complete multiple reflection \gls{qd} model}
Upon the first scattering event, all components produced -- both specular and diffuse -- behave as independent components and their remaining paths are traced accordingly.
We assume that every diffuse ray closely follows the path of the main cursor and further generates $N_{pre}+N_{post}$ diffuse \glspl{mpc} at each bounce.
The total number of \glspl{mpc} generated by a single deterministic rays at the $n$-th reflection will thus be $N_{MPC} \sim \qty(N_{pre} + 1 + N_{post})^n$.

\paragraph*{Reduced multiple reflection \gls{qd} model}
\begin{algorithm}[t] \caption{Reduced Multiple Reflection QD Generator} \label{alg:multipleReflectionQdGenerator}
\begin{algorithmic}[1]
\small
\Function{GetMpcsMultipleReflection}{Cursor, MaterialList, MaterialLibrary}
    \State CursorOutput $\leftarrow$ Cursor

    \vspace*{1ex}

    \For{Material $\in$ MaterialList}
        \State OtherMaterialsList $\leftarrow$ MaterialList $\setminus$ \{Material\}
        \State PreCursors, PostCursors $\leftarrow \varnothing$

        \vspace*{1ex}

        \State CurrentPreCursors, CursorOutput, CurrentPostCursors $\leftarrow$ \Call{GetMpcsFirstReflection}{CursorOutput, Material}

        \vspace*{1ex}

        \State PreCursors $\leftarrow$ Concatenate(PreCursors, \Call{OtherMaterialsReflLoss}{CurrentPreCursors, OtherMaterialsList, MaterialLibrary})
        \State PostCursors $\leftarrow$ Concatenate(PostCursors, \Call{OtherMaterialsReflLoss}{CurrentPostCursors, OtherMaterialsList, MaterialLibrary})
    \EndFor

    \vspace*{1ex}
    
    \Return PreCursors, CursorOutput, PostCursors
\EndFunction

\vspace*{2ex}

\Function{OtherMaterialsReflLoss}{Cursors, OtherMaterialsList, MaterialLibrary}
    \For{Cursor $\in$ Cursors}
        \For{Material $\in$ OtherMaterialsList}
            \State $RL \leftarrow \ricianDistrib{s_{RL, Material}}{\sigma_{RL, Material}}$
            \State Cursor.$PG$ $\leftarrow$ Cursor.$PG$ + $(RL - \mu_{RL, Material})$
        \EndFor
    \EndFor

    \Return Cursors
\EndFunction

\end{algorithmic}
\end{algorithm}

In order to reduce the exponential complexity of the complete model, the reduced model neglects diffuse rays beyond a first order given their multiplicatively high attenuation.
Instead, only diffuse rays generated directly by the deterministic ray are taken into account, each generated with the \gls{qd} parameters relative to the impinging reflecting surface.
Moreover, we assume that every diffuse component closely follows the main cursor, thus reflecting on the same reflectors (see \cref{alg:multipleReflectionQdGenerator}).
Consequently, every reflector produces $N_{pre} + N_{post}$ diffuse components, thus yielding a maximum of $N_{MPC} \sim n (N_{pre} + N_{post}) + 1$, including the deterministic ray and possible rays discarded during their generation (see \cref{sub:first_order_reflections}).


\section{Comparison with Measurements}
\label{sec:comparison_with_measurements}
Given the structure of this \gls{qd} model, every material must have a set of parameters for it to be appropriately simulated.
It follows that given the CAD file of an environment, every surface must be associated with a material with all the necessary simulation parameters taken, for example, from a material library.

\begin{figure}[t]
  \centering
  \setlength\belowcaptionskip{-.6cm}
  \setlength\fwidth{0.5\columnwidth} 
  \setlength\fheight{0.8\columnwidth}
  \input{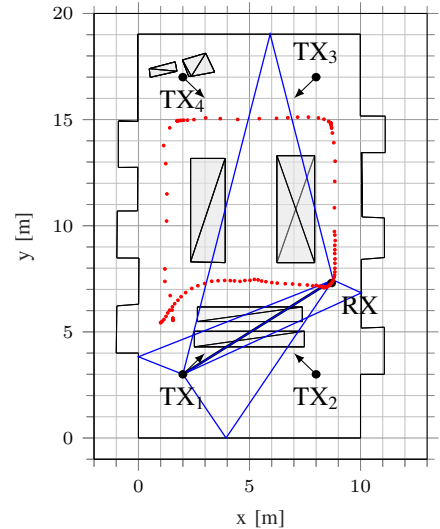}
  \caption{CAD model of NIST's lecture room.
  The 108~\gls{rx} positions from the measurement traces are shown in red. As an example, the direct and first reflection rays generated with the \gls{rt} for TX\textsubscript{1} and the specific \gls{rx} position are shown in black and blue, respectively.}
  \label{fig:lecture_room}
\end{figure}

We report in the following tables examples of material libraries from NIST's Lecture Room, reformulating the mean and variance provided per material~\cite{lai2019methodology} into the $s$ and $\sigma$ parameters needed to generate the random parameters of the model.
Measured data were taken from different \gls{tx} positions pointing towards the center of the room, where a mobile \gls{rx} sounder moved around the tables.
Specifically, as shown in~\cref{fig:lecture_room}, considering the bottom-left corner as the origin $(x_0, y_0, z_0) = (0,0,0)$, TX\textsubscript{1} is positioned in $(2, 3, 2.5)$~m, TX\textsubscript{2} in $(8,3,2.5)$~m, TX\textsubscript{3} in $(8,17,2.5)$~m, TX\textsubscript{4} in $(2,17,2.5)$~m, and the \gls{rx} performs a loop around the table.

\begin{table*}[t]
    \centering
    \renewcommand*{\arraystretch}{0.8}
    \caption{NIST's Lecture Room material library.}
    \label{tab:material_library_combined_materials}
    \scriptsize
    \begin{tabular}{lc|cccccc}
    \toprule
    && Left Wall (TX\textsubscript{2}) & Bottom Wall (TX\textsubscript{3}) & Right Wall (TX\textsubscript{1}) & Top Wall (TX\textsubscript{1}) & Tables (TX\textsubscript{1}) & Ceiling (TX\textsubscript{1}) \\ 
    \midrule 
    \multirow{2}{*}{$K_{dB} \sim \mathcal{R}(s,\sigma)$} & $(s_{K_{pre}}, \sigma_{K_{pre}})$   & (5.1196, 1.7485) & (1.4809, 2.1325) & (0, 0) & (0.5913, 4.5206) & (0, 0) & (3.6167, 7.2715) \\ 
    & $(s_{K_{post}}, \sigma_{K_{post}})$      & (6.2208, 3.5421) & (7.1809, 2.5325) & (0.2641, 3.1699) & (0.33, 3.7213) & (3.7738, 1.8748) & (7.1103, 2.2712) \\ 
    \midrule
    \multirow{2}{*}{$\gamma \sim \mathcal{R}(s,\sigma)$} & $(s_{\gamma_{pre}}, \sigma_{\gamma_{pre}})$          & (0.6742, 0.9992) & (0.9006, 0.2325) & (0, 0) & (0.0094, 0.2285) & (0, 0) & (0.9595, 0.901) \\ 
    & $(s_{\gamma_{post}}, \sigma_{\gamma_{post}})$         & (0.0658, 1.2034) & (0.6881, 0.3566) & (0.0412, 0.8648) & (0.0792, 1.1572) & (0.53, 0.4837) & (0.0717, 1.2794) \\
    \midrule
    \multirow{2}{*}{$\sigma_{s} \sim \mathcal{R}(s,\sigma)$} & $(s_{\sigma_{s,pre}}, \sigma_{\sigma_{s,pre}})$        & (0.0119, 0.3087) & (0.5553, 0.129) & (0, 0) & (0.243, 0.273) & (0, 0) & (0.2122, 0.0935) \\ 
    & $(s_{\sigma_{s,post}}, \sigma_{\sigma_{s,post}})$       & (0.4144, 0.1507) & (0.26, 0.1003) & (0.6367, 0.3209) & (0.201, 0.1901) & (0.3309, 0.4614) & (0.7679, 0.2484) \\ 
    \midrule
    \multirow{2}{*}{$\lambda \sim \mathcal{R}(s,\sigma)$} & $(s_{\lambda_{pre}}, \sigma_{\lambda_{pre}})$         & (0.9775, 0.3449) & (0.9172, 0.2241) & (0, 0) & (0.619, 1.1299) & (0, 0) & (0.8119, 0.2421) \\ 
    & $(s_{\lambda_{post}}, \sigma_{\lambda_{post}})$        & (0.8153, 0.6948) & (1.4106, 0.5832) & (0.9879, 0.4235) & (0.8655, 0.3762) & (0.8099, 0.076) & (0.7785, 0.1426) \\ 
    \midrule
    \multirow{2}{*}{$\sigma_{\alpha} \sim \mathcal{R}(s,\sigma)$} & $(s_{\sigma_{\alpha, az}}, \sigma_{\sigma_{\alpha, az}})$       & (0.1016, 2.2504) & (1.9426, 1.5726) & (3.2889, 1.3202) & (2.117, 2.1206) & (1.6594, 3.1974) & (1.9829, 0.9094) \\ 
    & $(s_{\sigma_{\alpha, el}}, \sigma_{\sigma_{\alpha, el}})$       & (2.9947, 1.6613) & (2.6946, 1.3948) & (3.2812, 1.8865) & (2.741, 1.7964) & (4.0345, 2.6859) & (2.696, 1.1135) \\  
    \midrule
    \multirow{2}{*}{$RL \sim \mathcal{R}(s,\sigma)$} & $(s_{RL}, \sigma_{RL})$        & (9.8412, 3.4424) & (8.5025, 4.2343) & (10.1562, 3.5164) & (6.7238, 5.9352) & (5.2106, 3.4013) & (6.5833, 2.1943) \\ 
    & $\mu_{RL}$      & 10.7 & 9.84 & 10.8 & 9.27 & 6.58 & 6.9 \\ 
    \bottomrule 
\end{tabular}
\end{table*}

Given that the channel sounder's \gls{tx} had a limited angular \gls{fov}, it was possible to characterize different surfaces, e.g., different walls by varying the \gls{tx} positions during the measurement campaign.
The model parameters per position have been reformatted accordingly in \cref{tab:material_library_combined_materials}.
Please note that, given the geometry of the room and the limited \gls{fov}, it was not possible to properly characterize some materials, such as the floor~\cite{lai2019methodology}.
For these materials, since no characterization was available, the pre/post cursors were not generated and the statistics for the reflection loss were taken from the ceiling instead.

\begin{figure*}[t!]
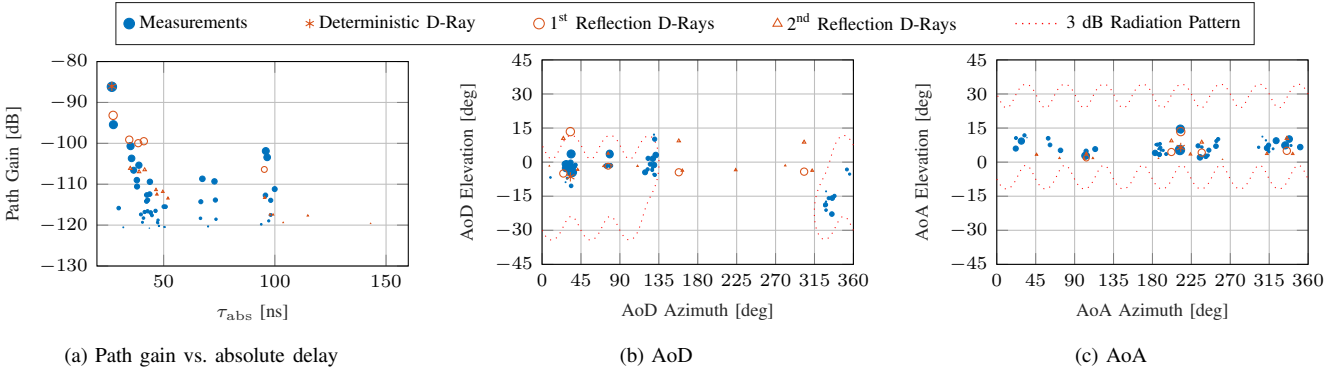

  \hspace{0.08\textwidth}
  \begin{subfigure}[b]{0.3\textwidth}
    \centering
    \setlength\fwidth{0.8\textwidth} 
    \setlength\fheight{0.8\textwidth}
%
%
\definecolor{mycolor1}{rgb}{0.00000,0.44700,0.74100}%
\definecolor{mycolor2}{rgb}{0.85000,0.32500,0.09800}%
\begin{tikzpicture}
\pgfplotsset{every tick label/.append style={font=\scriptsize}}

\begin{axis}[%
width=0,
height=0,
at={(0\fwidth,0\fheight)},
scale only axis,
xmin=0,
xmax=0,
xtick={},
ymin=0,
ymax=0,
ytick={},
axis background/.style={fill=white},
legend style={legend cell align=center, align=center, draw=white!15!black, font=\scriptsize, at={(0, 0)}, anchor=center, /tikz/every even column/.append style={column sep=2em}},
legend columns=5,
]
\addplot[scatter, only marks, mark=*, color=mycolor1, mark options={}, scatter/use mapped color={mark options={}, draw=mycolor1, fill=mycolor1}, visualization depends on={\thisrow{size} \as \perpointmarksize}, scatter/@pre marker code/.append style={/tikz/mark size=\perpointmarksize}] table[row sep=crcr]{%
x y size\\
0 0 0\\
};
\addlegendentry{Measurements}

\addplot[only marks, mark=asterisk, mark options={}, mark size=1.7695pt, draw=mycolor2] table[row sep=crcr]{%
x y\\
-1 -1\\
};
\addlegendentry{Deterministic D-Ray}

\addplot[scatter, only marks, mark=o, color=mycolor2, mark options={}, scatter/use mapped color=mycolor2, visualization depends on={\thisrow{size} \as \perpointmarksize}, scatter/@pre marker code/.append style={/tikz/mark size=\perpointmarksize}] table[row sep=crcr]{%
x y size\\
0 0 0\\
};
\addlegendentry{1\textsuperscript{st} Reflection D-Rays}

\addplot[scatter, only marks, mark=triangle, color=mycolor2, mark options={}, scatter/use mapped color=mycolor2, visualization depends on={\thisrow{size} \as \perpointmarksize}, scatter/@pre marker code/.append style={/tikz/mark size=\perpointmarksize}] table[row sep=crcr]{%
x y size\\
0 0 0\\
};
\addlegendentry{2\textsuperscript{nd} Reflection D-Rays}

\addplot [color=red, dotted, line width=0.5pt]
  table[row sep=crcr]{%
-1 -1\\
};
\addlegendentry{3~dB Radiation Pattern}

\end{axis}
\end{tikzpicture}%
  \end{subfigure}
  \\
  \hfill
  \begin{subfigure}[b]{0.3\textwidth}
    \centering
    \setlength\fwidth{0.8\textwidth} 
    \setlength\fheight{0.5\textwidth}
%
%
\definecolor{mycolor1}{rgb}{0.00000,0.44700,0.74100}%
\definecolor{mycolor2}{rgb}{0.85000,0.32500,0.09800}%
\definecolor{mycolor3}{rgb}{0.92900,0.69400,0.12500}%
\begin{tikzpicture}
\pgfplotsset{every tick label/.append style={font=\scriptsize}}

\begin{axis}[%
width=0.951\fwidth,
height=\fheight,
at={(0\fwidth,0\fheight)},
scale only axis,
xmin=20,
xmax=160,
xlabel style={font=\scriptsize\color{white!15!black}},
xlabel={$\tau_{\rm abs}$ [ns]},
ymin=-130,
ymax=-80,
ytick={-130, -120,..., -80},
ylabel style={font=\scriptsize\color{white!15!black}},
ylabel={Path Gain [dB]},
axis background/.style={fill=white},
xmajorgrids,
ymajorgrids,
]
\addplot[scatter, only marks, mark=*, color=mycolor1, mark options={}, scatter/use mapped color={mark options={}, draw=mycolor1, fill=mycolor1}, visualization depends on={\thisrow{size} \as \perpointmarksize}, scatter/@pre marker code/.append style={/tikz/mark size=\perpointmarksize}] table[row sep=crcr]{%
x	y	size\\
50.2950206502024	-120.442090390393	0.186365507593407\\
69.974856986181	-120.32128311338	0.213653857250528\\
47.8860570763082	-120.202228286662	0.237498403423384\\
46.3444745206577	-116.514568806257	0.624190627194372\\
47.4110162656404	-119.48449265624	0.348221820088938\\
43.5877782889284	-120.757305373138	0.0790569415042095\\
73.5285357532389	-118.559051601594	0.452635798176662\\
40.9289134375837	-118.275500775747	0.480104329978024\\
39.7774118760149	-117.387417123235	0.557445394946737\\
50.4988996697371	-120.484882754618	0.175685749558481\\
66.7196930928524	-114.27970383212	0.769123663081354\\
97.2446019965417	-118.961603769999	0.410494184750013\\
47.1876628306848	-119.318277164074	0.36915753600941\\
66.5587147241543	-118.331813765128	0.474775584020377\\
40.4652075462517	-119.310236614905	0.370140259850689\\
93.8788697131418	-119.7989408199	0.304705795117352\\
98.1372749786753	-113.935677920566	0.789073254481533\\
41.4401629465372	-116.792418508711	0.603745043565781\\
29.7856142923843	-115.837691422952	0.67139781500611\\
98.1206139271223	-117.521671196511	0.546456291390291\\
44.1326914053366	-116.987990239818	0.588928403836802\\
47.4744882050914	-118.777935287652	0.430234050556449\\
44.7303738082074	-117.601651001446	0.539803377888883\\
50.0838006316575	-115.487677614488	0.694551272725041\\
42.3183387284338	-114.165313144016	0.775813952672563\\
73.2645790223246	-113.864615897689	0.793131516356227\\
50.8226841676487	-115.497867068578	0.693888157239526\\
99.9380262885956	-111.188414398231	0.933205780002411\\
31.856387300584	-120.543201360955	0.159987334407633\\
42.7558808902946	-116.606900062457	0.617471536726829\\
43.7111455968835	-112.416943258163	0.871702919306559\\
42.404147761563	-112.628661139328	0.86066000120802\\
95.7749431309742	-112.707749161479	0.856498351139828\\
67.4111208120383	-108.676079615831	1.04779856506106\\
72.8470388799169	-109.305230053981	1.02031036614201\\
38.0361594459077	-110.565765876212	0.962878039679682\\
36.5217586073822	-106.579402772845	1.13460686609513\\
43.8173983001653	-109.405219393163	1.01587323817717\\
37.9459675902851	-108.993699280974	1.03401278954408\\
44.2143590732856	-116.875846469318	0.597469428989859\\
42.7840642304627	-113.85902224888	0.793450081294596\\
38.8639152707469	-105.344490377162	1.18275804395182\\
96.5411651859942	-103.424204832494	1.25396538440338\\
95.9289104866321	-101.885913832242	1.30821467262737\\
35.5740600236839	-103.681117498413	1.24467471509854\\
27.4037057176696	-95.4344064618975	1.51471747511325\\
35.117773035263	-100.677141387335	1.34931372692168\\
26.7277846807928	-86.1725402973838	1.76953383691864\\
};

\addplot[only marks, mark=asterisk, mark options={}, mark size=1.7695pt, draw=mycolor2] table[row sep=crcr]{%
x	y\\
26.708	-86.074\\
};

\addplot[scatter, only marks, mark=o, color=mycolor2, mark options={}, scatter/use mapped color=mycolor2, visualization depends on={\thisrow{size} \as \perpointmarksize}, scatter/@pre marker code/.append style={/tikz/mark size=\perpointmarksize}] table[row sep=crcr]{%
x	y	size\\
38.5641	-99.9649	1.3550329599617\\
95.2685	-106.39	1.1122328784981\\
34.5367	-99.1068	1.3842399951933\\
41.0852	-99.4849	1.37144731262289\\
27.285	-93.1597	1.57181607533499\\
};

\addplot[scatter, only marks, mark=triangle, color=mycolor2, mark options={}, scatter/use mapped color=mycolor2, visualization depends on={\thisrow{size} \as \perpointmarksize}, scatter/@pre marker code/.append style={/tikz/mark size=\perpointmarksize}] table[row sep=crcr]{%
x	y	size\\
46.9611	-112.476	0.545537878499805\\
49.6172	-111.824	0.569762560611408\\
38.9659	-106.955	0.725529227376786\\
99.1995	-117.441	0.303100992739467\\
97.806	-117.419	0.304601141348111\\
143.054	-119.591	0.052704627669473\\
51.9355	-113.45	0.507198809491002\\
114.812	-117.681	0.286225452855481\\
46.556	-111.371	0.58600421852677\\
34.9848	-106.119	0.749022772145796\\
41.4627	-106.464	0.739417936899064\\
95.4318	-113.305	0.513087898758965\\
103.746	-119.361	0.110944113413229\\
};

\end{axis}
\end{tikzpicture}%
    \caption{Path gain vs. absolute delay}
    \label{fig:pgTau_meas}
  \end{subfigure}
  \hfill
  \begin{subfigure}[b]{0.3\textwidth}
    \centering
    \setlength\fwidth{0.8\textwidth} 
    \setlength\fheight{0.5\textwidth}
    \input{img/tx1_t48_s01_refl2_qdoff_aod.tex}
    \caption{\gls{aod}}
    \label{fig:aod_meas}
  \end{subfigure}
  \hfill
  \begin{subfigure}[b]{0.3\textwidth}
     \centering
    \setlength\fwidth{0.8\textwidth} 
    \setlength\fheight{0.5\textwidth}
    \input{img/tx1_t48_s01_refl2_qdoff_aoa.tex}
    \caption{\gls{aoa}}
    \label{fig:aoa_meas}
  \end{subfigure}
  \hfill
  \setlength\belowcaptionskip{-.3cm}
  \caption{Example of comparison between measurements and ray-tracer, based on the channel between TX\textsubscript{1} and the \gls{rx} shown in \cref{fig:lecture_room} in the bottom left corner of the loop.
  In \subref{fig:pgTau_meas}, $\tau_{\rm abs}$ represents the absolute delay of each ray.
  \subref{fig:aod_meas} and \subref{fig:aoa_meas} show the 3~dB radiation patterns of the channel sounders described in~\cite{lai2019methodology} approximated with Gaussian beams.
  In fact, \glspl{mpc} outside of these regions are not detected in the measurements.}
  \label{fig:measurements_vs_rt}
\end{figure*}

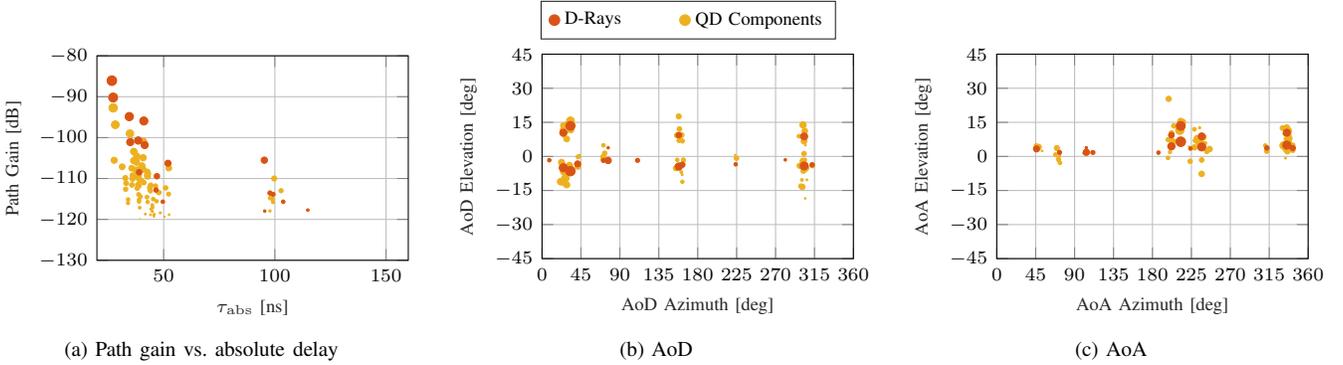
\begin{figure*}[t!]
  \hspace{0.35\textwidth}
  \begin{subfigure}[b]{0.3\textwidth}
    \centering
    \setlength\fwidth{0.8\textwidth} 
    \setlength\fheight{0.8\textwidth}
%
%
\definecolor{mycolor1}{rgb}{0.00000,0.44700,0.74100}%
\definecolor{mycolor2}{rgb}{0.85000,0.32500,0.09800}%
\definecolor{mycolor3}{rgb}{0.92900,0.69400,0.12500}%
\begin{tikzpicture}
\pgfplotsset{every tick label/.append style={font=\scriptsize}}

\begin{axis}[%
width=0,
height=0,
at={(0\fwidth,0\fheight)},
scale only axis,
xmin=0,
xmax=0,
xtick={},
ymin=0,
ymax=0,
ytick={},
axis background/.style={fill=white},
legend style={legend cell align=center, align=center, draw=white!15!black, font=\scriptsize, at={(0, 0)}, anchor=center, /tikz/every even column/.append style={column sep=2em}},
legend columns=5,
]
\addplot[scatter, only marks, mark=*, color=mycolor2, mark options={}, scatter/use mapped color={mark options={}, draw=mycolor1, fill=mycolor1}, visualization depends on={\thisrow{size} \as \perpointmarksize}, scatter/@pre marker code/.append style={/tikz/mark size=\perpointmarksize}] table[row sep=crcr]{%
x y size\\
0 0 0\\
};
\addlegendentry{D-Rays}

\addplot[scatter, only marks, mark=*, color=mycolor3, mark options={}, scatter/use mapped color={mark options={}, draw=mycolor1, fill=mycolor1}, visualization depends on={\thisrow{size} \as \perpointmarksize}, scatter/@pre marker code/.append style={/tikz/mark size=\perpointmarksize}] table[row sep=crcr]{%
x y size\\
0 0 0\\
};
\addlegendentry{QD Components}

\end{axis}
\end{tikzpicture}%
  \end{subfigure}
  \\
  \hfill
  \begin{subfigure}[b]{0.3\textwidth}
    \centering
    \setlength\fwidth{0.8\textwidth} 
    \setlength\fheight{0.5\textwidth}
%
%
\definecolor{mycolor1}{rgb}{0.00000,0.44700,0.74100}%
\definecolor{mycolor2}{rgb}{0.85000,0.32500,0.09800}%
\definecolor{mycolor3}{rgb}{0.92900,0.69400,0.12500}%
\begin{tikzpicture}
\pgfplotsset{every tick label/.append style={font=\scriptsize}}

\begin{axis}[%
width=0.951\fwidth,
height=\fheight,
at={(0\fwidth,0\fheight)},
scale only axis,
xmin=20,
xmax=160,
xlabel style={font=\scriptsize\color{white!15!black}},
xlabel={$\tau_{\rm abs}$ [ns]},
ymin=-130,
ymax=-80,
ytick={-130, -120,..., -80},
ylabel style={font=\scriptsize\color{white!15!black}},
ylabel={Path Gain [dB]},
axis background/.style={fill=white},
xmajorgrids,
ymajorgrids,
]
\addplot[scatter, only marks, mark=*, color=mycolor3, mark options={}, scatter/use mapped color={mark options={}, draw=mycolor3, fill=mycolor3}, visualization depends on={\thisrow{size} \as \perpointmarksize}, scatter/@pre marker code/.append style={/tikz/mark size=\perpointmarksize}] table[row sep=crcr]{%
x	y	size\\
26.708	-86.074	1.76953383691864\\
38.5641	-100.712011771896	1.33282533904169\\
95.2685	-105.524563989086	1.15368792431532\\
34.5367	-94.8639744234086	1.52239635211838\\
41.0852	-95.9303457491281	1.48962822841045\\
27.285	-90.2012753299853	1.6580857936498\\
46.9611	-109.440160330777	0.984167848075861\\
49.6172	-115.676413836673	0.62560977774909\\
38.9659	-108.474778204584	1.02856096289243\\
99.1995	-113.847532870605	0.748772562941483\\
97.806	-113.556732219214	0.766534743698506\\
51.9355	-106.274764549479	1.12319220761401\\
114.812	-117.722659899242	0.449440240507953\\
46.556	-112.814510178023	0.810106258434948\\
34.9848	-101.096082095508	1.31942245431818\\
41.4627	-101.823077684499	1.29367239525255\\
95.4318	-117.970684096219	0.423131844853228\\
103.746	-115.68421057822	0.625032770105601\\
36.0667609037783	-109.241405169907	0.993469783592609\\
34.0198211505525	-115.137950081638	0.664247925724785\\
33.9856333903646	-116.014821052164	0.600055154558714\\
39.2306573442492	-117.142982818246	0.505617204307567\\
39.3182975347403	-119.837572593111	0.0790569415042095\\
34.84363251641	-99.0021003853615	1.39092949777309\\
36.4902567540524	-105.629922332373	1.14945393388408\\
36.5652886991128	-103.408473059803	1.23565828954252\\
37.5854480289944	-105.555475368148	1.15244731443243\\
37.8456456884108	-106.058041486016	1.13208640079602\\
37.9146956832423	-104.535434623944	1.19270493519178\\
38.9145577960903	-111.779785478444	0.867203251139326\\
41.7700087745542	-118.703934546631	0.333428088778742\\
40.8193187987811	-100.995646606881	1.32294045856462\\
39.1153948282924	-107.729519914921	1.06156267462154\\
38.9131902497703	-115.259907169822	0.655696212618376\\
42.5398789387744	-109.288895749367	0.991255114838383\\
42.6316343665695	-108.371005595999	1.03321941847385\\
42.6873540203075	-109.408622584816	0.985649705004669\\
42.8879389457001	-109.526432110707	0.980102768364868\\
42.962848189628	-108.821564590724	1.0128379493445\\
43.802537330736	-112.405446052281	0.833146584957328\\
43.8248318564438	-112.126557169437	0.848496255824115\\
44.0204966446785	-113.029802936343	0.797712760808516\\
44.1780606394763	-111.663012626823	0.873412521912703\\
44.884933736052	-116.517863734869	0.559916856507978\\
45.035727436105	-117.088207232994	0.510606043762373\\
45.4599342571868	-118.959161752447	0.295891266464052\\
45.5900003807334	-118.985931421255	0.29167441597065\\
45.6540959203055	-119.319717276967	0.232766602199869\\
45.8127240995101	-118.731004009376	0.32964961126876\\
27.2932408677337	-92.7240278985652	1.58611292102855\\
27.7459789126186	-105.545093668082	1.15286412670685\\
28.2642282228124	-96.8536067056174	1.46066406068665\\
31.3478213836138	-107.13268664239	1.08726971692045\\
32.7667097448075	-109.891563260228	0.962707945025504\\
32.9148134423664	-112.623049909842	0.820970660973183\\
33.1560566298358	-111.142334813542	0.90057812643\\
38.1400432085439	-109.249604856185	0.993087753374798\\
38.9573826754277	-115.637531116747	0.62847943371099\\
46.8086401580518	-115.561808012089	0.634030739531113\\
47.7345128950889	-113.444793214327	0.773263260128527\\
47.9716598593921	-113.437682297708	0.773688711946264\\
48.4643207185383	-118.409096706692	0.37210632372349\\
50.881607143581	-112.286519446828	0.839726460318524\\
50.3590731415247	-119.350757945058	0.226511170887876\\
38.7615431930822	-113.914140313614	0.744644528879714\\
38.3979697985227	-111.513201902587	0.881314484534407\\
38.0978501039719	-112.362083544548	0.835551712961346\\
38.8612615103328	-110.126679752649	0.951338684772605\\
39.6051922580524	-114.453419269917	0.710339573950359\\
39.6668418090354	-110.419485372876	0.936986958408054\\
41.0052319628136	-110.978496008884	0.908958262635553\\
97.816093370053	-117.923465757631	0.428264951860678\\
97.877356074424	-114.78916341716	0.6881187296448\\
98.9077020459814	-114.924527379211	0.678954140064499\\
99.1922988336882	-115.752269562216	0.619973176167644\\
99.6938341004481	-109.998252139258	0.957565641925392\\
102.774674638454	-112.984490658109	0.800337143643944\\
52.4229320377598	-118.860112989863	0.310997004123831\\
52.2150683151489	-113.809049937541	0.75114722470033\\
52.3122013494708	-107.452383726859	1.07357615237795\\
46.3213033335009	-111.875411954273	0.862085108793718\\
45.3250243774545	-115.042924858645	0.67083559955698\\
35.7444128326468	-111.555685562694	0.879080843672494\\
37.5068219964479	-107.886941995421	1.05467770101948\\
38.8725309792003	-109.241649749911	0.993458390532006\\
42.7626621072547	-115.746172740839	0.620428104403749\\
44.3601756808514	-117.47429467682	0.474324851956444\\
35.6135453240429	-107.486351611726	1.07211092931479\\
36.0238140988117	-110.06767509017	0.95420463821601\\
40.860802625485	-107.787364938823	1.05903798066011\\
40.7560241198555	-104.922838639338	1.17757748786113\\
40.2821337693637	-106.872339907365	1.0982950086732\\
42.5208095114067	-114.560513295464	0.70332793419172\\
44.0583246740729	-118.886056001504	0.307112310885031\\
44.9162238439859	-118.049875439601	0.414380228694878\\
};

\addplot[scatter, only marks, mark=*, color=mycolor2, mark options={}, scatter/use mapped color={mark options={}, draw=mycolor2, fill=mycolor2}, visualization depends on={\thisrow{size} \as \perpointmarksize}, scatter/@pre marker code/.append style={/tikz/mark size=\perpointmarksize}] table[row sep=crcr]{%
x	y	size\\
26.708	-86.074	1.76953383691864\\
38.5641	-100.712011771896	1.33282533904169\\
95.2685	-105.524563989086	1.15368792431532\\
34.5367	-94.8639744234086	1.52239635211838\\
41.0852	-95.9303457491281	1.48962822841045\\
27.285	-90.2012753299853	1.6580857936498\\
46.9611	-109.440160330777	0.984167848075861\\
49.6172	-115.676413836673	0.62560977774909\\
38.9659	-108.474778204584	1.02856096289243\\
99.1995	-113.847532870605	0.748772562941483\\
97.806	-113.556732219214	0.766534743698506\\
51.9355	-106.274764549479	1.12319220761401\\
114.812	-117.722659899242	0.449440240507953\\
46.556	-112.814510178023	0.810106258434948\\
34.9848	-101.096082095508	1.31942245431818\\
41.4627	-101.823077684499	1.29367239525255\\
95.4318	-117.970684096219	0.423131844853228\\
103.746	-115.68421057822	0.625032770105601\\
};

\end{axis}
\end{tikzpicture}%
    \caption{Path gain vs. absolute delay}
    \label{fig:pgTau_qd}
  \end{subfigure}
  \hfill
  \begin{subfigure}[b]{0.3\textwidth}
    \centering
    \setlength\fwidth{0.8\textwidth} 
    \setlength\fheight{0.5\textwidth}
%
%
\definecolor{mycolor3}{rgb}{0.00000,0.44700,0.74100}%
\definecolor{mycolor2}{rgb}{0.85000,0.32500,0.09800}%
\definecolor{mycolor3}{rgb}{0.92900,0.69400,0.12500}%
\begin{tikzpicture}
\pgfplotsset{every tick label/.append style={font=\scriptsize}}

\begin{axis}[%
width=0.951\fwidth,
height=\fheight,
at={(0\fwidth,0\fheight)},
scale only axis,
xmin=0,
xmax=360,
xtick={0, 45, ..., 360},
xlabel style={font=\scriptsize\color{white!15!black}},
xlabel={AoD Azimuth [deg]},
ymin=-45,
ymax=45,
ytick={-45, -30, ..., 45},
ylabel style={font=\scriptsize\color{white!15!black}},
ylabel={AoD Elevation [deg]},
axis background/.style={fill=white},
xmajorgrids,
ymajorgrids,
]
\addplot[scatter, only marks, mark=*, color=mycolor3, mark options={}, scatter/use mapped color={mark options={}, draw=mycolor3, fill=mycolor3}, visualization depends on={\thisrow{size} \as \perpointmarksize}, scatter/@pre marker code/.append style={/tikz/mark size=\perpointmarksize}] table[row sep=crcr]{%
x	y	size\\
32.7153	-6.4539	1.76953383691864\\
158.095	-4.4648	1.33282533904169\\
76.4285	-1.8058	1.15368792431532\\
24.638	-4.9867	1.52239635211838\\
303.02	-4.19029999999999	1.48962822841045\\
32.7153	13.4313	1.6580857936498\\
162.178	-3.6652	0.984167848075861\\
223.925	-3.4688	0.62560977774909\\
158.095	9.36060000000001	1.02856096289243\\
110.405	-1.7342	0.748772562941483\\
70.9992	-1.7589	0.766534743698506\\
41.1561	-3.3138	1.12319220761401\\
281.226	-1.4983	0.449440240507953\\
312.31	-3.6972	0.810106258434948\\
24.638	10.4371	1.31942245431818\\
303.02	8.7923	1.29367239525255\\
76.4285	3.8079	0.423131844853228\\
8.39228	-1.65819999999999	0.625032770105601\\
154.913800913806	-5.18383620883971	0.993469783592609\\
156.309235611757	-1.50344790927861	0.664247925724785\\
163.23175996345	-1.40875660484495	0.600055154558714\\
160.918605309551	-6.87352066730084	0.505617204307567\\
157.876133803497	-4.23517374950438	0.0790569415042095\\
26.9359402293419	-9.60829784918469	1.39092949777309\\
22.5371858090192	-2.64663217226533	1.14945393388408\\
21.4812261034745	-11.1362828942487	1.23565828954252\\
28.4695247018964	-12.4874189624439	1.15244731443243\\
24.7161176134366	-6.70408499821549	1.13208640079602\\
31.6980758159158	-3.98922027312786	1.19270493519178\\
24.0654004123466	-11.7026447531189	0.867203251139326\\
29.8234230157401	-4.29217562376937	0.333428088778742\\
304.43653369246	-5.04523561373747	1.32294045856462\\
301.608896561939	-1.9906799047728	1.06156267462154\\
303.91377178731	-1.51097699931177	0.655696212618376\\
303.88835662511	-3.82240939476189	0.991255114838383\\
300.865053313669	-13.4198318847803	1.03321941847385\\
297.902355108133	-4.10772612196065	0.985649705004669\\
302.173450189646	-6.30152750822685	0.980102768364868\\
303.908520666546	-4.19405455332041	1.0128379493445\\
302.20361000382	-3.47409623835595	0.833146584957328\\
301.700557814292	-3.37844954145181	0.848496255824115\\
299.077125873284	-13.0503467225829	0.797712760808516\\
302.900576950697	1.19352600666809	0.873412521912703\\
302.192821620147	-3.10252462934562	0.559916856507978\\
300.450155501373	-10.3525112690542	0.510606043762373\\
304.197025145206	-18.5125943343647	0.295891266464052\\
305.452422521178	-5.38212242765485	0.29167441597065\\
301.856924652343	-3.68340047803439	0.232766602199869\\
306.15452904074	-10.3033063793248	0.32964961126876\\
32.3271400518274	12.5078984226884	1.58611292102855\\
27.402568326482	13.7012463961349	1.15286412670685\\
32.8290963051769	15.6618714125471	1.46066406068665\\
30.4631035403357	14.141131646394	1.08726971692045\\
35.6127344312691	13.6754152314904	0.962707945025504\\
28.6811424632472	12.6625476332923	0.820970660973183\\
33.7320307824328	11.4346327660646	0.90057812643\\
29.0828262201432	11.6576957454149	0.993087753374798\\
30.2893984737302	13.3019261567901	0.62847943371099\\
162.224341463951	-11.1579218178469	0.634030739531113\\
163.537813986882	-1.6664531567709	0.773263260128527\\
162.286800756359	6.77072533973835	0.773688711946264\\
161.999627759589	-8.00186550449905	0.37210632372349\\
224.827550441357	-0.79875933884523	0.839726460318524\\
223.173211916264	0.679330072256633	0.226511170887876\\
158.224048477615	6.97853452744495	0.744644528879714\\
158.876464521436	12.0292603528252	0.881314484534407\\
155.703515188781	9.2193604095408	0.835551712961346\\
158.060532115255	17.6079663625176	0.951338684772605\\
155.645883338623	9.44554018098839	0.710339573950359\\
159.596961046948	7.40281544347067	0.936986958408054\\
161.36354645135	9.41568899551255	0.908958262635553\\
71.6090181710609	0.890103726530626	0.428264951860678\\
70.232704437088	-1.23966559191005	0.6881187296448\\
70.9461288808634	4.88646827824333	0.678954140064499\\
71.7702604717162	1.46207407748776	0.619973176167644\\
70.8050659193769	-0.216255202706009	0.957565641925392\\
72.6635337495522	1.42763332993216	0.800337143643944\\
43.2696914966637	0.575826025336724	0.310997004123831\\
41.7410894364879	-0.192491335633804	0.75114722470033\\
41.3794299567536	-4.42694132960163	1.07357615237795\\
312.524545733575	-4.02331243220723	0.862085108793718\\
312.350233870346	-3.87647906488797	0.67083559955698\\
24.1115158865951	13.6731056356176	0.879080843672494\\
25.1015170015358	11.5711415171228	1.05467770101948\\
24.36884119726	11.8754599637787	0.993458390532006\\
24.7528917509769	10.0724560087244	0.620428104403749\\
24.7806918340806	8.99707056705451	0.474324851956444\\
28.4985054135262	7.67875776157547	1.07211092931479\\
26.0090098443253	12.9377030720752	0.95420463821601\\
303.770253499602	11.1826759679729	1.05903798066011\\
301.732870267293	13.8675679498055	1.17757748786113\\
302.917402509815	6.71160505383493	1.0982950086732\\
297.102502424507	8.88148201608156	0.70332793419172\\
304.590875152585	4.94314773515408	0.307112310885031\\
299.973204853286	8.80833432825114	0.414380228694878\\
};

\addplot[scatter, only marks, mark=*, color=mycolor2, mark options={}, scatter/use mapped color={mark options={}, draw=mycolor2, fill=mycolor2}, visualization depends on={\thisrow{size} \as \perpointmarksize}, scatter/@pre marker code/.append style={/tikz/mark size=\perpointmarksize}] table[row sep=crcr]{%
x	y	size\\
32.7153	-6.4539	1.76953383691864\\
158.095	-4.4648	1.33282533904169\\
76.4285	-1.8058	1.15368792431532\\
24.638	-4.9867	1.52239635211838\\
303.02	-4.19029999999999	1.48962822841045\\
32.7153	13.4313	1.6580857936498\\
162.178	-3.6652	0.984167848075861\\
223.925	-3.4688	0.62560977774909\\
158.095	9.36060000000001	1.02856096289243\\
110.405	-1.7342	0.748772562941483\\
70.9992	-1.7589	0.766534743698506\\
41.1561	-3.3138	1.12319220761401\\
281.226	-1.4983	0.449440240507953\\
312.31	-3.6972	0.810106258434948\\
24.638	10.4371	1.31942245431818\\
303.02	8.7923	1.29367239525255\\
76.4285	3.8079	0.423131844853228\\
8.39228	-1.65819999999999	0.625032770105601\\
};

\end{axis}
\end{tikzpicture}%
    \caption{\gls{aod}}
    \label{fig:aod_qd}
  \end{subfigure}
  \hfill
  \begin{subfigure}[b]{0.3\textwidth}
     \centering
    \setlength\fwidth{0.8\textwidth} 
    \setlength\fheight{0.5\textwidth}
%
%
\definecolor{mycolor3}{rgb}{0.00000,0.44700,0.74100}%
\definecolor{mycolor2}{rgb}{0.85000,0.32500,0.09800}%
\definecolor{mycolor3}{rgb}{0.92900,0.69400,0.12500}%
\begin{tikzpicture}
\pgfplotsset{every tick label/.append style={font=\scriptsize}}

\begin{axis}[%
width=0.951\fwidth,
height=\fheight,
at={(0\fwidth,0\fheight)},
scale only axis,
xmin=0,
xmax=360,
xtick={0, 45, ..., 360},
xlabel style={font=\scriptsize\color{white!15!black}},
xlabel={AoA Azimuth [deg]},
ymin=-45,
ymax=45,
ytick={-45, -30, ..., 45},
ylabel style={font=\scriptsize\color{white!15!black}},
ylabel={AoA Elevation [deg]},
axis background/.style={fill=white},
xmajorgrids,
ymajorgrids,
]
\addplot[scatter, only marks, mark=*, color=mycolor3, mark options={}, scatter/use mapped color={mark options={}, draw=mycolor3, fill=mycolor3}, visualization depends on={\thisrow{size} \as \perpointmarksize}, scatter/@pre marker code/.append style={/tikz/mark size=\perpointmarksize}] table[row sep=crcr]{%
x	y	size\\
212.715	6.4539	1.76953383691864\\
201.905	4.4648	1.33282533904169\\
103.549	1.8058	1.15368792431532\\
335.362	4.9867	1.52239635211838\\
236.98	4.19029999999999	1.48962822841045\\
212.715	13.4313	1.6580857936498\\
342.178	3.6652	0.984167848075861\\
223.925	3.4688	0.62560977774909\\
201.905	9.36060000000001	1.02856096289243\\
111.127	1.7342	0.748772562941483\\
72.5348	1.7589	0.766534743698506\\
46.0046	3.3138	1.12319220761401\\
101.203	1.4983	0.449440240507953\\
312.31	3.6972	0.810106258434948\\
335.362	10.4371	1.31942245431818\\
236.98	8.7923	1.29367239525255\\
103.549	3.8079	0.423131844853228\\
186.983	1.65819999999999	0.625032770105601\\
198.759532608951	25.335724280074	0.993469783592609\\
197.974853405731	-0.663299407606615	0.664247925724785\\
200.279692741805	1.91827166711386	0.600055154558714\\
202.624916676872	5.39815425843004	0.505617204307567\\
204.478932493028	4.84069170102333	0.0790569415042095\\
331.558604987593	4.78733711535219	1.39092949777309\\
333.970301588784	5.40994039792982	1.14945393388408\\
337.13375044875	11.2785927732408	1.23565828954252\\
334.319616648954	3.28599650861905	1.15244731443243\\
336.112761916379	2.37855929370195	1.13208640079602\\
337.185239703317	7.25736568824412	1.19270493519178\\
340.233990440843	5.1600656807135	0.867203251139326\\
333.782696415552	-0.780254035867742	0.333428088778742\\
234.018942187198	6.29649750732038	1.32294045856462\\
246.007596243053	3.23093768994468	1.06156267462154\\
240.746626098657	5.77909615499355	0.655696212618376\\
236.422194696175	3.7082317125559	0.991255114838383\\
236.889379862542	-7.70510675153876	1.03321941847385\\
238.069766536266	4.08248019964552	0.985649705004669\\
228.475289995196	3.98844780377934	0.980102768364868\\
237.553249566612	8.63603862716766	1.0128379493445\\
242.807013650887	2.15751161229394	0.833146584957328\\
229.697188650856	6.97161580468752	0.848496255824115\\
238.176714748026	5.7906842778884	0.797712760808516\\
236.89656826901	-1.58158537680877	0.873412521912703\\
236.928474344594	-1.72660518835501	0.559916856507978\\
236.855800513087	3.83056168616622	0.510606043762373\\
229.516011278565	3.29631984037742	0.295891266464052\\
235.347809014744	3.21929041381729	0.29167441597065\\
236.264245858774	3.47466558214772	0.232766602199869\\
237.086603947508	5.7194602998362	0.32964961126876\\
211.644995067532	11.5332312766093	1.58611292102855\\
212.641551561279	15.4547610669965	1.15286412670685\\
212.552373894178	13.4571866030121	1.46066406068665\\
214.992464891396	15.1074581559936	1.08726971692045\\
212.863578828413	15.1707290643704	0.962707945025504\\
211.197165401341	14.1314159624293	0.820970660973183\\
213.87310336013	12.2287641162815	0.90057812643\\
212.610822732785	12.0874693683381	0.993087753374798\\
214.346750955566	13.826627803641	0.62847943371099\\
342.703460193043	5.08836865181419	0.634030739531113\\
342.078399761274	4.55887675586089	0.773263260128527\\
342.414899632436	2.60448150293075	0.773688711946264\\
332.268519605262	5.89567715386403	0.37210632372349\\
223.640137193854	-2.14518567158457	0.839726460318524\\
222.243143549453	4.37066046721027	0.226511170887876\\
197.992698362095	7.375285168883	0.744644528879714\\
201.786856559746	7.00893135409004	0.881314484534407\\
204.664228752956	13.4330830458496	0.835551712961346\\
200.903380925136	7.27416225864644	0.951338684772605\\
201.941387391936	10.4161456397639	0.710339573950359\\
204.194230034655	6.60605984686839	0.936986958408054\\
202.176977747502	10.2160598984765	0.908958262635553\\
72.3858098197845	0.29583082564892	0.428264951860678\\
67.7821726273983	0.795937676522314	0.6881187296448\\
73.0486445008182	-2.84229906857826	0.678954140064499\\
71.6383085393759	-1.74019004886115	0.619973176167644\\
69.504704366478	3.66596191663976	0.957565641925392\\
71.314961940915	-1.11614059553085	0.800337143643944\\
52.3768957949034	2.40735445151245	0.310997004123831\\
45.6084960972942	4.70592687123231	0.75114722470033\\
48.2662716360924	4.11328801802297	1.07357615237795\\
312.1351582204	2.32685755532809	0.862085108793718\\
310.54392065109	4.77269658852273	0.67083559955698\\
330.341929760214	12.4583124147235	0.879080843672494\\
335.163616185271	12.7077360805046	1.05467770101948\\
338.365744059424	7.97618277331243	0.993458390532006\\
332.732277918475	9.94214184810484	0.620428104403749\\
336.107297902154	9.59652372787305	0.474324851956444\\
335.512581977317	7.71885797994652	1.07211092931479\\
337.305628085485	9.41782482761894	0.95420463821601\\
238.863267931502	8.51032400780986	1.05903798066011\\
236.67652148608	8.44876494256295	1.17757748786113\\
231.712558582848	7.99727202383858	1.0982950086732\\
228.578358353145	11.9277746376128	0.70332793419172\\
237.197693396483	9.02800240124911	0.307112310885031\\
234.550268971936	12.5880811475112	0.414380228694878\\
};

\addplot[scatter, only marks, mark=*, color=mycolor2, mark options={}, scatter/use mapped color={mark options={}, draw=mycolor2, fill=mycolor2}, visualization depends on={\thisrow{size} \as \perpointmarksize}, scatter/@pre marker code/.append style={/tikz/mark size=\perpointmarksize}] table[row sep=crcr]{%
x	y	size\\
212.715	6.4539	1.76953383691864\\
201.905	4.4648	1.33282533904169\\
103.549	1.8058	1.15368792431532\\
335.362	4.9867	1.52239635211838\\
236.98	4.19029999999999	1.48962822841045\\
212.715	13.4313	1.6580857936498\\
342.178	3.6652	0.984167848075861\\
223.925	3.4688	0.62560977774909\\
201.905	9.36060000000001	1.02856096289243\\
111.127	1.7342	0.748772562941483\\
72.5348	1.7589	0.766534743698506\\
46.0046	3.3138	1.12319220761401\\
101.203	1.4983	0.449440240507953\\
312.31	3.6972	0.810106258434948\\
335.362	10.4371	1.31942245431818\\
236.98	8.7923	1.29367239525255\\
103.549	3.8079	0.423131844853228\\
186.983	1.65819999999999	0.625032770105601\\
};

\end{axis}
\end{tikzpicture}%
    \caption{\gls{aoa}}
    \label{fig:aoa_qd}
  \end{subfigure}
  \hfill
  \setlength\belowcaptionskip{-.3cm}
  \caption{Reduced multiple reflection \gls{qd} model applied to \gls{rt}-based channel traces with up to 2\textsuperscript{nd} order reflections.
  Rays with path gain below -120~dB are not shown, to more closely resemble the dynamic range of the channel sounder.}
  \label{fig:rt_vs_qd}
\end{figure*}

\cref{fig:measurements_vs_rt} shows an example of measured channel compared to the deterministic ray-traced channel for the scenario of \cref{fig:lecture_room}.
As can be seen, the direct ray is correctly identified both in the power-delay domain and in the angles domains, while other rays only partially resemble the measurements.
This is due to (i) the approximated CAD model which may be missing some relevant reflectors and (ii) inaccuracies in the measurements.

While delays shown in \cref{fig:pgTau_meas} are in good accord between measurements and \gls{rt} simulation, path gains are less precise, due to the random reflection losses experienced by the rays.
Notice also that the \gls{tx} only has antennas towards the front (as shown by the antenna pattern in \cref{fig:aod_meas}), thus, rays predicted by the \gls{rt} to depart with an azimuth angle between 135\si{\degree} and 315\si{\degree} were not part of the real measurements.
Most of all, though, it is easily noticeable that there exist clusters of rays well defined in the joined path gain, delay, \gls{aod}, \gls{aoa} domain, and are missing, instead in the channel generated by the \gls{rt}.
Such clusters do not arise from higher order reflections (not shown here), but rather from diffuse \glspl{mpc}, thus highlighting the need for a valid diffuse \gls{qd} model.

\begin{figure*}[t!]
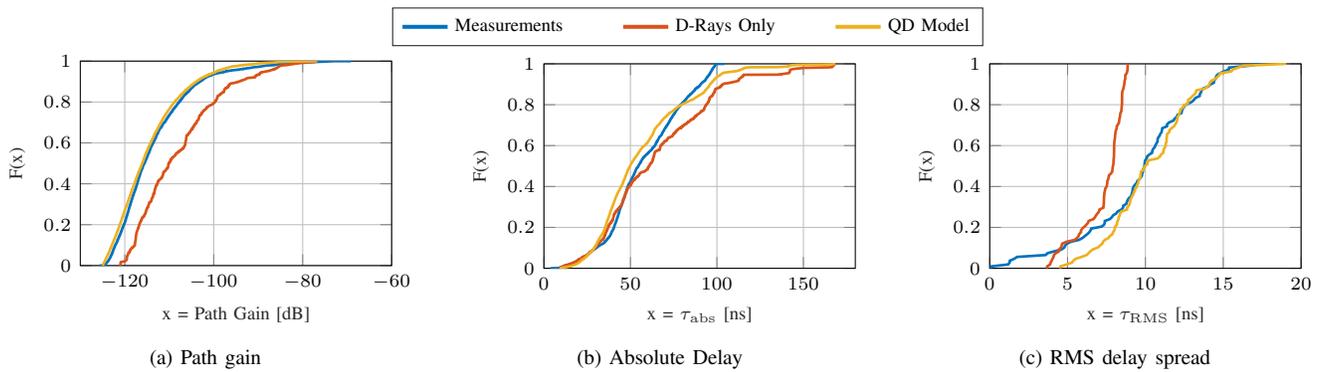

  \hspace{0.28\textwidth}
  \begin{subfigure}[b]{0.3\textwidth}
    \centering
    \setlength\fwidth{0.8\textwidth} 
    \setlength\fheight{0.8\textwidth}
%
%
\definecolor{mycolor1}{rgb}{0.00000,0.44700,0.74100}%
\definecolor{mycolor2}{rgb}{0.85000,0.32500,0.09800}%
\definecolor{mycolor3}{rgb}{0.92900,0.69400,0.12500}%
\begin{tikzpicture}
\pgfplotsset{every tick label/.append style={font=\scriptsize}}

\begin{axis}[%
width=0,
height=0,
at={(0\fwidth,0\fheight)},
scale only axis,
xmin=0,
xmax=0,
xtick={},
ymin=0,
ymax=0,
ytick={},
axis background/.style={fill=white},
legend style={legend cell align=center, align=center, draw=white!15!black, font=\scriptsize, at={(0, 0)}, anchor=center, /tikz/every even column/.append style={column sep=2em}},
legend columns=5,
]
\addplot [color=mycolor1, line width=1.5pt]
  table[row sep=crcr]{%
0  0\\
};
\addlegendentry{Measurements}

\addplot [color=mycolor2, line width=1.5pt]
  table[row sep=crcr]{%
0  0\\
};
\addlegendentry{D-Rays Only}

\addplot [color=mycolor3, line width=1.5pt]
  table[row sep=crcr]{%
0  1\\
};
\addlegendentry{QD Model}

\end{axis}
\end{tikzpicture}%
  \end{subfigure}
  \\
  \hfill
  \begin{subfigure}[b]{0.3\textwidth}
    \centering
    \setlength\fwidth{0.8\textwidth} 
    \setlength\fheight{0.5\textwidth}
    \input{img/pg_cdf.tex}
    \caption{Path gain}
    \label{fig:pg_cdf}
  \end{subfigure}
  \hfill
  \begin{subfigure}[b]{0.3\textwidth}
    \centering
    \setlength\fwidth{0.8\textwidth} 
    \setlength\fheight{0.5\textwidth}
    \input{img/delay_cdf.tex}
    \caption{Absolute Delay}
    \label{fig:delay_cdf}
  \end{subfigure}
  \hfill
  \begin{subfigure}[b]{0.3\textwidth}
     \centering
    \setlength\fwidth{0.8\textwidth} 
    \setlength\fheight{0.5\textwidth}
    \input{img/tauRms_cdf.tex}
    \caption{RMS delay spread}
    \label{fig:tauRms_cdf}
  \end{subfigure}
  \hfill
  \setlength\belowcaptionskip{-.3cm}
  \caption{Comparison between \acrshortpl{cdf} of \gls{mpc} path gain, absolute delay, and RMS delay spread with and without \gls{qd} model with respect to the measurements.}
  \label{fig:cdfs}
\end{figure*}

\cref{fig:rt_vs_qd,fig:cdfs} show how the proposed \gls{qd} model enhances the realism of a purely deterministic channel, making it significantly more similar to the measured one.
Specifically, \cref{fig:rt_vs_qd} reports an example of a specific channel instance, based on the CAD model shown in \cref{fig:lecture_room} and for the same \gls{tx}/\gls{rx} locations of \cref{fig:measurements_vs_rt}. With respect to the \gls{rt} specular reflections from \cref{fig:measurements_vs_rt}, the deterministic rays (in orange), which are generated up to second order reflections, also include a random reflection loss component in the path gain. The diffuse rays added to the model are plotted in blue, with sizes proportional to the respective path gain.
By comparing \cref{fig:rt_vs_qd} with \cref{fig:measurements_vs_rt}, it is clear that the \glspl{dray} alone are not able to fully model the complexity of a real channel, and that the proposed \gls{qd} model can instead play an important role to this regard.
In fact, empirically, rays are parts of clusters with small variations in the angular and delay domains, and large variations in the power gain domain.

Furthermore, the effects of the added rays are clearly shown in \cref{fig:cdfs}, which plots the \glspl{cdf} of the path gain (\cref{fig:pg_cdf}), the absolute delay (\cref{fig:delay_cdf}), and the RMS delay spread (\cref{fig:tauRms_cdf}), similar to the RMS angle spread shown in~\cite{lai2019methodology}, for the multipath components of the scenarios.
The CDFs show the combined statistics of the mmWave channel between TX\textsubscript{1} and 108~\gls{rx} positions shown in red in \cref{fig:lecture_room}).
Notably, it is clear how the delays and path gains generated with the proposed \gls{qd} model are significantly closer to the real measurements with respect to purely deterministic rays alone, with \gls{cdf} fit improvements from 73~\% to 86~\% (i.e., \gls{ks} test improvements of 0.13) for the path gain, from 86~\% to 89~\% (i.e., \gls{ks} test improvements of 0.03) for the absolute delay, and from 33~\% to 87~\% (i.e., \gls{ks} test improvements of 0.54) for the RMS delay spread.

\section{Conclusions}
\label{sec:conclusions}
Performance evaluation is a fundamental part of the design of 5G mmWave networks.
To that end, an accurate channel model allows researchers to generate reliable simulation results, that can qualitatively and quantitatively describe what can be expected when using real devices.
In this paper, we introduce a mathematical formulation for a class of mmWave channels, i.e., the \gls{qd} models, that can closely simulate the propagation of rays in a specific environment.
We provided a step-by-step tutorial on how such models can be implemented, including the parameters and random distributions obtained from a NIST measurement campaign~\cite{lai2019methodology}.
We then compared the results that can be obtained with an open source implementation of the model with the real measurement traces, showing improvements in the \gls{ks} test for path gain (0.131), delay (0.03), and RMS delay spread (0.537).

As future work, we will further extend the \gls{qd} model with material libraries from other measurement campaigns, and study methods to reduce the computational complexity involved in the ray and channel matrices generation, as in~\cite{lecci2020simplified}.

\bibliographystyle{IEEEtran}
\bibliography{bibl}

\end{document}